\documentclass[twocolumn,floats,showpacs,superscriptaddress,pre]{revtex4}
\usepackage{amsmath,amssymb,bm}
\usepackage{graphics}
\usepackage{epsfig}
\usepackage{graphicx}

\input{tcilatex}
\begin{document}

\title{Does the Universe have its own mass?}
\author{Natalia Gorobey}
\affiliation{Peter the Great Saint Petersburg Polytechnic University, Polytekhnicheskaya
29, 195251, St. Petersburg, Russia}
\author{Alexander Lukyanenko}
\email{alex.lukyan@mail.ru}
\affiliation{Peter the Great Saint Petersburg Polytechnic University, Polytekhnicheskaya
29, 195251, St. Petersburg, Russia}
\author{A. V. Goltsev}
\affiliation{Ioffe Physical- Technical Institute, Polytekhnicheskaya 26, 195251, St.
Petersburg, Russia}

\begin{abstract}
Within the framework of the previously proposed formulation of the quantum
theory of gravity in terms of world histories, it was suggested that the
universe has its own mass. This quantity is analogous to the mass of a
particle in relativistic mechanics. The mass of the universe is a
distribution of non-zero values of gravitational constraints, which arises
and changes in time as a consequence of the initial conditions for
fundamental dynamic variables. A formulation of the Euclidean quantum theory
of gravity is also proposed to determine the initial state, which can be the
source of the universe's own mass. Being unrelated to ordinary matter, the
distribution of its own mass affects the geometry of space and forms a
dedicated frame of reference. The existence of selected reference systems is
taken into account by the corresponding modification of the system of
quantum gravitational links. A variant of such a modification of the
Wheeler-De Witt equation is the operator representation of gravitational
constraints, which, together with the state of the universe, determines the
parameters of the reference system in the form of a distribution of the
spinor field on a spatial section.
\end{abstract}

\maketitle

%\author{Natalia Gorobey$^{1}$, Alexander Lukyanenko$^{1,\ast }$, and A.V.
%Goltsev$^{2}$}
%\email{$^{\ast }$alex.lukyan@mail.ru}
%\affiliation{$^{1}$Peter the Great Saint Petersburg Polytechnic University,
%Polytekhnicheskaya 29, 195251, St. Petersburg, Russia\\
%$^{2}$Ioffe Physical-Technical Institute, Polytekhnicheskaya 26, 195251, St.
%Petersburg, Russia}

%\date{\today }

%\pacs{}

%\begin{multicols}{2}
%\narrowtext

%%%%%%%%%%%%%%%%%%%%%%%%%%%%%%%%%%%%%%%%%%%%%%%%%%%%%%%%%%%%%%%%%%%%%%%

%%\bigskip

\section{\textbf{INTRODUCTION}}

Speaking about the mass of the universe in this work, we mean the analogy
with the mass of a particle in relativistic mechanics, which is included in
the well-known relativistic relation between energy and momentum (we assume
the speed of light to be unity):
\begin{equation}
H\equiv p_{\mu }p^{\mu }=-p_{0}^{2}+p_{i}^{2}+m^{2}=0.  \label{1}
\end{equation}%
This equation has a direct analogy with the constraint equations in the
canonical representation of the theory of gravity of Arnowitt, Deser, and
Mizner (ADM) \cite{MTW,ADM}. The latter will be the basis of our
consideration. The question of what is the dynamic nature of the particle
mass arose after Dirac and Fock introduced the proper time of the particle
as an independent dynamic variable \cite{Dir,Fock}. Subsequently, St%
\~{o}ckelberg, Feynman, and Schwinger introduced proper time into quantum
electrodynamics \cite{St,Fe,Schw} and considered mass as a
dynamic variable conjugate to proper time (see also \cite{Alv}). In this
paper, we ask ourselves the question: should the mass in equation Eq.(\ref{1}%
) be considered an independent constant, or is it a consequence of the
conditions in the source of the particle that were imposed at its birth? It
is clear that the answer to this question should be sought in quantum
theory. Another aspect closely related to the concepts of proper time and
mass is the proper covariant formulation of quantum theory. The generally
accepted formulation of the quantum theory of gravity (QTG) is based on the
quantum version of gravitational constraints - the Wheeler-DeWitt (WDW)
equations \cite{Wheel},\cite{DeWitt}. It implements the results of canonical
analysis and Dirac's proposals on quantization of covariant theories \cite%
{Dir1}. This formulation rules out the possibility of a non-zero self-mass
of the universe. Another approach based on the invariant definition of the
Batalin-Fradkin-Vilkovysky (BFV) functional integral \cite{Frad},\cite{Bat}
gives an object that can be called a propagator (Green's function of ADM
constraints) in the state space, which includes additional integration over
proper time (see about BFW theorem in the case of a relativistic particle
\cite{Gov}). Written in the Euclidean representation, the invariant
functional integral over all Riemannian $4D$ metrics with one spatial
boundary (and "south pole") defines the no-boundary Hartle-Hawking wave
function of the universe - the proposed non-singular solution of WDW \cite%
{HH}. After integration over the Euclidean proper time, there are also no
sources for the appearance of the universe's own mass.

In this paper, we propose an alternative formulation of the covariant QTG,
in which there is an explicit time parameter that describes the evolution of
the universe in terms of trajectories in the configuration space
(superspace). In the new formulation, the ADM connections, while remaining
canonical generators of dynamics, can be nonzero. Their numerical value is
determined by the initial conditions at the time of the birth of the
universe. The modified covariant quantum dynamics, after the transition to
the Euclidean form, gives a variant of the description of the "subpolar
region" or the cosmological vacuum that determines the initial state of the
universe. We allow for a natural violation of covariance in this region, so
that the initial state can be the source of the universe's own mass. Note
that the distribution and motion of this self-mass in space forms a
dedicated frame of reference and \textquotedblleft
spontaneously\textquotedblright\ breaks the covariance. This circumstance
can be considered as one of the examples of spontaneous vacuum symmetry
breaking.

However, the description of the evolution of the universe in terms of the
external time parameter, although covariant in form and independent of
parametrization, is a view "from outside". Since all observers are "inside"
the universe, the problem of determining the internal time parameter remains
relevant. In this paper, we propose a variant of modification of the ADM
links, which also includes restrictions on the parameters of the frame of
reference. This variant is based on the Witten identity \cite{Witt,Fadd}, 
which was first obtained in connection with the proof of the
gravitational field energy positivity theorem.

In the next section, we briefly formulate the standard approach to the
covariant quantum theory of gravity. In the second section, a modification
of this quantum theory is proposed, in which a non-zero self-mass of the
universe is allowed, and a variant of the initial state of the universe in
which this mass arises is also proposed. In the third section, a
modification of the WDW system of equations is proposed, which includes the
parameters of the selected reference system. The conditional principle of
the extremum of the energy of space is also formulated, which can be used as
the basis for determining the internal time of the universe.

\section{COVARIANT QUANTUM THEORY OF GRAVITY}

Here we briefly formulate the generally accepted approach to the quantum
theory of gravity. Consideration begins with the classical action of the
general theory of relativity (GR),
\begin{equation}
I\left[ g,\varphi \right] =\frac{1}{16\pi G}\int R\sqrt{-g}d^{4}x+I_{m}\left[
g,\varphi \right] ,  \label{2}
\end{equation}%
where the second term is the action functional of matter fields. We will
keep in mind the need to add appropriate boundary contributions to obtain a
canonical representation and subsequent quantization (see \cite{Haw}).
Variation of the action with respect to the metric tensor $g_{\mu \nu }$
gives the Einstein equations
\begin{equation}
R^{\mu \nu }-\frac{1}{2}g^{\mu \nu }R=16\pi GT^{\mu \nu },  \label{3}
\end{equation}
\begin{equation}
T^{\mu \nu }\equiv \frac{\delta I_{m}}{\delta g_{\mu \nu }}.  \label{4}
\end{equation}%
The transition to the canonical representation is carried out by following the
ADM \cite{MTW}. Using the $3+1$ splitting of the $4D$ metric
\begin{equation}
ds^{2}=g_{ik}\left( dx^{i}+N^{i}dt\right) \left( dx^{k}+N^{k}dt\right)
-\left( Ndt\right) ^{2},  \label{5}
\end{equation}%
where $i,k=1,2,3$, we represent the density of the Lagrange function of the
Hilbert-Einstein action in the form
\begin{equation}
16\pi G\mathit{L}=\mathbf{\pi }^{ik}\frac{\partial g_{ik}}{\partial t}-N%
\mathit{H}-N_{i}\mathit{H}^{i},  \label{6}
\end{equation}%
where

\begin{equation}
N_{i}=g_{ik}N^{k},  \label{7}
\end{equation}%
and

\begin{equation}
\mathbf{\pi }^{ik}=\sqrt{\det g_{lm}}\left( g^{ik}Tr\mathbf{K-}K^{ik}\right)
,  \label{8}
\end{equation}%
have the meaning of canonical momenta conjugate to the components of the $3D$
metric $g_{ik}$, in which

\begin{equation}
K_{ik}=\frac{1}{2N}\left( N_{i\left\vert k\right. }+N_{k\left\vert i\right.
}-\frac{\partial g_{ik}}{\partial t}\right) .  \label{9}
\end{equation}%
The density of the Hamilton function is a linear combination of
gravitational constraints,

\begin{equation}
\mathit{H}\equiv \frac{1}{\sqrt{\det g_{lm}}}\left( Tr\mathbf{\pi }^{2}-%
\frac{1}{2}\left( Tr\mathbf{\pi }\right) ^{2}\right) ,  \label{10}
\end{equation}

\begin{equation}
\mathit{H}^{i}=-2\mathbf{\pi }_{\left\vert k\right. }^{ik}.  \label{11}
\end{equation}%
Here and below, for simplicity, we do not include matter fields. The
components of the $4D$ metrics $N,N_{i}$ are called succession and shift
functions and play the role of Lagrange multipliers in the canonical form of
the ADM action. A variation of the action with respect to these factors
gives the classical constraint equations:

\begin{equation}
\mathit{H}=\mathit{H}^{i}=0.  \label{12}
\end{equation}%
It is these equalities that will be the subject of discussion and possible
modification in this paper. They are part of the Einstein equations Eq.(\ref%
{3}) and a necessary consequence of the general covariance of the theory.
Although in the canonical formalism based on the action of the ADM, the
dynamic meaning of the components $N,N_{i}$ of the $4D$ metric differs from
the rest of the components, the Euler-Lagrange equations Eq.(\ref{12})
corresponding to them must be strictly fulfilled.

The quantization of the theory is carried out in the standard way by
replacing the canonical momenta with the corresponding Hermitian
differentiation operators on the space of wave functions $\psi \left(
g_{ik}\right) $,

\begin{equation}
\widehat{\pi }^{kl}\left( x_{m}\right) =\frac{\hbar }{i}\frac{\delta }{%
\delta g_{kl}\left( x_{m}\right) }.  \label{13}
\end{equation}%
Their substitution into Eqs.(\ref{10}),(\ref{11}) gives the constraint
operators $\widehat{\mathit{H}}$,$\widehat{\mathit{H}}^{i}$ (the problem of
ordering noncommuting factors is not discussed here). In the generally
accepted version of the covariant quantization theory, in accordance with
Dirac's proposal \cite{Dir1}, the classical constraint equations Eq.(\ref{13}%
) are replaced by additional conditions for the wave function. In the case
of the quantum theory of gravity, these additional conditions have the form
of the WDW equations,

\begin{equation}
\widehat{\mathit{H}}\psi \left( g_{ik}\right) =\widehat{\mathit{H}}^{i}\psi
\left( g_{ik}\right) =0.  \label{14}
\end{equation}%
Here we use parentheses to denote the functional dependence of the wave
function on the metric, assuming the subsequent introduction of a functional
on world histories with a time parameter. In the WDW equations Eq.(\ref{14}%
), there is no time parameter, which in the case of a closed universe
creates a difficult problem for the interpretation of the theory. The WDW
solutions form the set of admissible physical states of the universe in the
covariant quantum theory. The construction of a specific solution requires
additional conditions for the constraints Eq.(\ref{14}). One such solution,
the no-boundary wave function, was proposed by Hartle and Hawking \cite{HH}.

To approach the definition of the no-boundary wave function of the universe,
it is useful to refer to an analogy with the simplest covariant quantum
theory of a relativistic particle based on the Klein-Gordon (KG) equation:

\begin{equation}
\widehat{H}\psi \left( x\right) =\left( \hbar ^{2}\partial _{\mu }\partial
^{\mu }+m^{2}\right) \psi \left( x\right) =0.  \label{15}
\end{equation}%
In this case, the BFW invariant functional integral reduces to the
expression (see \cite{Gov})

\begin{equation}
G\left( x,x^{\prime }\right) =-i\int_{0}^{\infty }K\left( x,x^{\prime
},s\right) ds,  \label{16}
\end{equation}%
where $K\left( x,x^{\prime },s\right) $ is the solution of the parabolic
equation

\begin{equation}
i\hbar \frac{\partial K}{\partial s}=\widehat{H}K  \label{17}
\end{equation}%
with the corresponding initial condition (see also \cite{DeWitt1})

\begin{equation}
K\left( x,x^{\prime },0\right) =\delta \left( x-x^{\prime }\right) .
\label{18}
\end{equation}%
The function $G(x,x\prime )$ is the Feynman propagator of a particle is a
singular solution of the KG equation (for $x\rightarrow x\prime $).

Returning to the theory of gravity, we can also start with the universe
propagator in the form of a covariant functional integral over all
pseudo-Euclidean $4D$ geometries between two fixed spatial sections, which
is a singular solution of the WDW equations. In this context, the suggestion
of Hartl and Hawking is understandable - to obtain a non-singular solution
of the WDW equations by removing one of the boundary surfaces, which is
achieved by passing to imaginary time and integrating over all Riemannian $%
4D $ geometries with one fixed spatial section. In this form of covariant
quantum theory of gravity, there is no external parameter of temporal
evolution, and there is no place for the universe's own mass. The mass of
the universe can be introduced into quantum theory if time is returned there
as a parameter of evolution.

\section{OWN MASS OF THE UNIVERSE}

The interpretation of nonrelativistic quantum mechanics based on the KG
equation for a charged particle (for example, a pi meson) is based on the
theory of electric charge perturbations \cite{BD}. Here the particle and the
antiparticle correspond to positive- and negative-particle solutions of the
KG equation. There is also an interpretation of the solutions of the
parabolic wave equation with proper time Eq.(\ref{17}), in which sections of
world lines directed backward in time in Minkowski space are compared to
antiparticles \cite{St}. Here $K\left( x,x^{\prime },s\right) $ serves as
the kernel of the evolution operator for Eq.(\ref{17}).

Let's give this interpretation in terms of world lines another form, more
suitable for generalization. To this end, we start with the action of a
relativistic particle in a parametrized form (with an arbitrary parameter $%
\tau $),

\begin{equation}
I\left[ x\right] =\int_{0}^{1}\left( \frac{\overset{\cdot }{x}^{2}}{4N}%
-m^{2}N\right) d\tau ,  \label{19}
\end{equation}%
which has a clear analogy with the ADM representation of the
Hilbert-Einstein action. We write the classical equations of motion of a
free particle, which follow from (\ref{19}), in the form

\begin{equation}
\frac{d^{2}x^{\mu }}{ds^{2}}=0,  \label{20}
\end{equation}%
where the proper time parameter is explicitly introduced, according to

\begin{equation}
ds=Nd\tau .  \label{21}
\end{equation}%
Note that the introduction of the proper time as an evolution parameter in
Eq.(\ref{19}) makes the second term with the mass redundant in the particle
dynamics. It is essential for determining proper time using the additional
constraint equation

\begin{equation}
\frac{1}{4}\frac{dx^{\mu }dx_{\mu }}{ds^{2}}=-m^{2},  \label{22}
\end{equation}%
which is obtained by varying Eq.(\ref{19}) with respect to $N$. Let us
introduce the canonical momenta

\begin{equation}
p_{\mu }=\frac{1}{2}\overset{\cdot }{x}_{\mu },  \label{23}
\end{equation}%
and write the action Eq.(\ref{19}) in the canonical form

\begin{equation}
I\left[ x\right] =\int_{0}^{1}\left( p_{\mu }\overset{\cdot }{x}^{\mu
}-NH\right) d\tau ,  \label{24}
\end{equation}%
where $H$ is determined by relation Eq.(\ref{1}), an illogical form of the
ADM theory of gravity. Relativistic quantum mechanics is obtained by
replacing the 4-momentum of the particle by the differentiation operators

\begin{equation}
\widehat{p}_{\mu }=\frac{\hbar }{i}\frac{\partial }{\partial x^{\mu }},
\label{25}
\end{equation}%
substituting which into Eq.(\ref{1}) we obtain the KG equation. In this
quantum theory, the probability measure

\begin{equation}
\left( \frac{i\hbar }{2m}\right) \left( \overline{\psi }\frac{\partial \psi
}{\partial t}-\psi \frac{\partial \overline{\psi }}{\partial t}\right)
\label{26}
\end{equation}%
is sign indefinite in accordance with the interpretation of positive- and
negative-frequency solutions of the KG equation \cite{BD}.

Passing to the interpretation in terms of world lines, we will normalize the
solutions of Eq.(\ref{17}) $\psi (x,s)$ by the quadratic form

\begin{equation}
\left\langle \psi \right\vert \left. \psi \right\rangle =\int d^{4}x%
\overline{\psi }\left( x,s\right) \psi \left( x,s\right) .  \label{27}
\end{equation}%
Let us divide the time interval $[0,S]$ into small segments of length $%
\varepsilon $ by points $s_{n}=nS/N$, $n=1,2,\ldots ,N$, and approximate an
arbitrary world line $x^{\mu }=x^{\mu }\left( s\right) $ by a polyline with
vertices $x_{n}^{\mu }=x^{\mu }\left( s_{n}\right) $. Let us introduce the
multiplicative function of the polygonal vertices

\begin{equation}
\Psi \left( x_{n}\right) =\prod\limits_{n}\psi \left( x_{n},s_{n}\right) ,
\label{28}
\end{equation}%
where $\psi \left( x,s\right) $ the solution of equation Eq.(\ref{17}) on
the interval $[0,S]$. We define the norm of this function by the quadratic
form

\begin{equation}
\left\langle \Psi \right\vert \left. \Psi \right\rangle =\int
\prod\limits_{n}d^{4}x\overline{\Psi }\Psi .  \label{29}
\end{equation}%
Thus, the function $\Psi \left( x_{n}\right) $ determines the probability of
movement along some world line passing through the points of the polyline
with vertices x\_n, provided that the initial wave function $\psi _{0}\left(
x\right) $ is given. As such, we take a wave packet with a given initial
4-momentum $p_{0\mu }$:

\begin{equation}
\psi _{0}\left( x\right) =A\exp \left[ -\sum\limits_{\mu }\frac{\left(
x^{\mu }-x_{0}^{\mu }\right) ^{2}}{2\sigma _{\mu }^{2}}+\frac{i}{\hbar }%
p_{0\nu }x^{\nu }\right] ,  \label{30}
\end{equation}%
where we put

\begin{equation}
p_{0}^{2}=p_{0\mu }p_{0}^{\mu }=-m^{2}.  \label{31}
\end{equation}%
This wave packet obviously describes the state of a particle in a source
localized near the point $x_{0\mu }$ of the Minkowski space, which has
finite dimensions $\sigma _{\mu }^{2}$ (source coherence parameters). Using
function Eq.(\ref{28}), one can calculate the average acceleration values

\begin{eqnarray}
&&\left\langle \frac{d^{2}x^{\mu }\left( s\right) }{ds^{2}}\right\rangle
_{\Psi }  \notag \\
&=&\frac{\left\langle x^{\mu }\left( s_{n+1}\right) \right\rangle _{\psi
}-2\left\langle x^{\mu }\left( s_{n}\right) \right\rangle _{\psi
}+\left\langle x^{\mu }\left( s_{n-1}\right) \right\rangle _{\psi }}{%
\varepsilon ^{2}}  \notag \\
&=&0,  \label{32}
\end{eqnarray}%
and squared particle velocity in Minkowski space

\begin{eqnarray}
\left\langle \frac{dx^{\mu }dx_{\mu }}{ds^{2}}\right\rangle _{\Psi } &=&%
\frac{\left\langle x^{\mu }\left( s_{n+1}\right) x_{\mu }\left(
s_{n+1}\right) \right\rangle _{\psi }}{\varepsilon ^{2}}  \notag \\
&&-\frac{2\left\langle x^{\mu }\left( s_{n+1}\right) \right\rangle _{\psi
}\left\langle x_{\mu }\left( s_{n}\right) \right\rangle _{\psi }}{%
\varepsilon ^{2}}  \notag \\
+\frac{\left\langle x^{\mu }\left( s_{n}\right) x_{\mu }\left( s_{n}\right)
\right\rangle _{\psi }}{\varepsilon ^{2}} &\neq &0.  \label{33}
\end{eqnarray}%
In relation Eq.(\ref{32}), the Ehrenfest theorem \cite{Ehr} is formulated
for a relativistic particle, and the average in Eq.(\ref{33}) is determined
by the mass $m_{0}^{2}$ of the particle in the source (as well as the
coherence parameters $\sigma _{\mu }^{2}$ of the source). Thus, in quantum
theory, a particle has a mass that is entirely determined by the initial
state, if its kinematic mass $m^{2}$ in the KG equation is set equal to zero.

We will come to a new formulation of the RQM in terms of world lines of a
particle if we pass to the limit $\varepsilon \rightarrow 0$, in which the
broken line approximates an arbitrary world line $x^{\mu }=x^{\mu }\left(
s\right) $ arbitrarily exactly. In this limit, the function $\Psi \left(
x_{n}\right) $ turns into a wave functional $\Psi \left[ x\left( s\right) %
\right] $ on the space of particle world lines, and the Schr\"{o}dinger
equation Eq.(\ref{17}) is replaced by the quantum principle of least action
\cite{GorLukGol1} ,which is a secularequation for the action operator

\begin{eqnarray}
&&\widehat{I}\Psi  \notag \\
&\equiv &\int_{0}^{S}ds\left[ \frac{\widetilde{\hbar }}{i}\overset{\cdot }{x}%
^{\mu }\left( s\right) \frac{\delta \Psi }{\delta x^{\mu }\left( s\right) }+%
\widetilde{\hbar }^{2}\frac{\delta ^{2}\Psi }{\delta x^{\mu }\left( s\right)
\delta x_{\mu }\left( s\right) }\right]  \notag \\
&=&\Lambda \Psi ,  \label{34}
\end{eqnarray}%
where

\begin{equation}
\widetilde{\hbar }=\hbar \cdot \varepsilon ,  \label{35}
\end{equation}%
and the limit $\varepsilon \rightarrow 0$ is assumed. Based on this form of
RQM, we will now present the necessary modification of the QTG, which we
will also use to determine the initial state of the universe.

This modification will be based on the parabolic Schr\"{o}dinger equation
for the wave function of the universe $\psi \left( g_{ik},t\right) $

\begin{equation}
i\hbar \frac{\partial \psi }{\partial t}=\left( N\widehat{\mathit{H}}+N_{k}%
\widehat{\mathit{H}}^{k}\right) \psi ,  \label{36}
\end{equation}%
with coordinate time $t$, in which the following and shift functions are
arbitrary and fixed. This equation is analogous to Eq.(\ref{17}). Proceeding
further in the same way as in the case of a particle, we divide the time
interval on which the dynamics of the universe is considered into small
segments of length $\varepsilon $ and compose a multiplicative wave function
(in the limit $\varepsilon \rightarrow 0$, the wave functional)

\begin{equation}
\Psi \left( g_{ik}\left( x_{l},t_{n}\right) \right) =\prod\limits_{n}\psi
\left( g_{ik}\left( x_{l},t_{n}\right) \right) .  \label{37}
\end{equation}%
With this limit in mind, we define the generalized momentum operator on the
space of wave functionals

\begin{equation}
\widehat{\pi }^{kl}\left( x_{m},t\right) =\frac{\widetilde{\hbar }}{i}\frac{%
\delta }{\delta g_{kl}\left( x_{m},t\right) }.  \label{38}
\end{equation}%
Replacing the canonical impulses by operators in the canonical form of the
ADM action (we agree to place them on the right in all terms), we obtain the
action operator $\widehat{I}$. As in ordinary quantum mechanics, the secular
equation for this operator

\begin{eqnarray}
\widehat{I}\Psi &\equiv &\int d^{4}x\left[ \frac{\widetilde{\hbar }}{i}\frac{%
\partial g_{kl}\left( x_{m},t\right) }{\partial t}\frac{\delta }{\delta
g_{kl}\left( x_{m},t\right) }\right.  \notag \\
&&\left. -N\widehat{\mathit{H}}-N_{k}\widehat{\mathit{H}}^{k}\right] \Psi
\notag \\
&=&\Lambda \Psi ,  \label{39}
\end{eqnarray}%
where the eigenvalue is determined by the boundary values of the metric on
the initial and final spatial sections,

\begin{equation}
\frac{i}{\hbar }\Lambda =\ln \psi \left( g_{ik},T\right) -\ln \psi \left(
g_{ik},0\right) ,  \label{40}
\end{equation}%
($T$ is considered time interval) is equivalent to the Schr\"{o}dinger
equation Eq.(\ref{37}). We assume that the eigenfunctional $\Psi \left[ g%
\right] $ - the solution of this secular equation is an invariant of
transformations of the space-time coordinates that do not affect the
boundary surfaces. We emphasize that $\Psi $ is the world history functional
$g_{\alpha \beta }\left( x^{k},t\right) $, including the dependence of the
metric on time. As in the case of a relativistic particle, sections of
history are allowed, with backward movement in time, i.e. compression of
parts of the universe ($\det g_{ik}\rightarrow 0$) with the formation of
black holes. We define the invariant norm of the wave functional $\Psi \left[
g\left( x\right) \right] $ by the quadratic form

\begin{equation}
\left\langle \Psi \right\vert \left. \Psi \right\rangle =\int
\prod\limits_{x}J\left[ g\right] d^{10}g\overline{\Psi }\left[ g\right] \Psi %
\left[ g\right] ,  \label{41}
\end{equation}%
where $J[g]$is the corresponding element of the Faddeev-Popov invariant
measure \cite{FP}. Based on it the probabilistic interpretation of QTG in a
new formulation, we can calculate the average values of the Einstein
equations Eq.(\ref{2}) (taking into account at this stage also the matter
fields):

\begin{equation}
\left\langle R^{\mu \nu }-\frac{1}{2}g^{\mu \nu }R-16\pi GT^{\mu \nu
}\right\rangle _{\Psi }.  \label{43}
\end{equation}%
Now the question is whether these averages are equal to zero. Among them are
the mean values of the connections $\mathit{H}^{a}=\left( \mathit{H,H}%
^{i}\right) $, which in the classical theory are equal to zero as a
consequence of the extremum of the classical action with respect to the
lapse and shift functions. Now this extremum condition is absent, as well as
there is no functional integration over the lapse and shift functions, as
provided for in the invariant form of the BFV functional integral. Thus, in
the new formulation of quantum theory, the equality of constraints to zero
is not mandatory. For the remaining Einstein equations, we assume the
validity of the Ehrenfest theorem. In this case, the constraint algebra
determined by the commutation relations

\begin{equation}
\left\{ \mathit{H}^{a},\mathit{H}^{b}\right\} =C_{d}^{ab}\mathit{H}^{d},
\label{44}
\end{equation}%
retains its meaning in the new formulation. Here the structural "constants" $%
C_{d}^{ab}$ are functions of the $3D$ metric $g_{ik}$ \cite{ADM}, and
summation over indices also implies integration over spatial coordinates. It
follows from Eq.(\ref{43}) that the average values $\left\langle \mathit{H}%
^{a}\right\rangle $, which are scalar and vector densities in space, also
depend on time:

\begin{equation}
\frac{\partial }{\partial t}\left\langle \mathit{H}^{a}\right\rangle
=\left\langle C_{d}^{ab}N_{b}\mathit{H}^{d}\right\rangle .  \label{45}
\end{equation}%
Thus, the mean values $\left\langle \mathit{H}^{a}\right\rangle $ are always
and everywhere equal to zero if they are equal to zero at the beginning.
Whether this is so depends on the initial state of the universe.

One of the options for determining the initial state of the universe was
proposed in \cite{GorLukGol2}. In its construction, a new representation of
the QTG is used in terms of the quantum principle of least action Eq.(\ref%
{39}). To do this, we pass to the Euclidean form of action by Wick's
rotation of time in the complex plane, $t\rightarrow it$, with the
simultaneous transformation of the canonical momenta, $\mathbf{\pi }%
\rightarrow -i\mathbf{\pi }$. The Euclidean representation allows us to
formulate the quantum principle of least action for a $4D$ geometry with one
"spatial" section. We pay attention to the fact that in Riemannian geometry
the specificity of the time coordinate is completely lost and for Euclidean
quantization the generalized canonical form of De-Donder-Weil follows \cite%
{DeDon},\cite{Weil}. This is also allowed by the quantum principle of least
action, which for the initial state takes the form (see \cite{GorLukGol2})

\begin{eqnarray}
\widehat{I}_{E}\Psi _{0} &\equiv &\int d^{4}x\frac{\widetilde{\hbar }%
_{\alpha }}{i}\partial _{\alpha }g_{\beta \gamma }\left( x\right) \frac{%
\delta \Psi _{0}}{\delta _{\alpha }g_{\beta \gamma }\left( x\right) }  \notag
\\
&&-\mathcal{H}\left[ g_{\beta \gamma }\left( x\right) ,\frac{\delta }{\delta
_{\alpha }g_{\beta \gamma }\left( x\right) }\right] \Psi _{0}  \notag \\
&=&\Lambda _{0}\Psi _{0},  \label{46}
\end{eqnarray}%
where the integral is taken over a compact domain of a $4D$ dimensional
Riemannian space with one boundary. Here

\begin{equation}
\widetilde{\hbar }_{\alpha }=\hbar \cdot \varepsilon _{\alpha }  \label{47}
\end{equation}%
where $\varepsilon _{\alpha }$ are spatial lattice constants (see \cite%
{GorLukGol2}). The eigenvalue of the action $\Lambda _{0}$ now depends only
on the boundary values of the $4D$ metric and determines the initial state
of the universe at this boundary for the subsequent dynamics of the state in
time:

\begin{equation}
\psi _{0}\left( g_{ik}\right) =\exp \left[ \frac{i}{\hbar }\Lambda
_{0}\left( g_{ik}\right) \right] .  \label{48}
\end{equation}%
Note that the De-Donder-Weil canonical formalism as applied to a metric
field requires the fulfillment of an additional condition

\begin{equation}
\det g_{\alpha \beta }=const,  \label{49}
\end{equation}%
which violates the $4D$ covariance of this initial state theory. Thus, there
is no reason to attribute the initial state Eq.(\ref{48}) to the set of
solutions for WDW equations Eq.(\ref{14}). It can be assumed that the
subsequent evolution of the universe with such an initial state will include
additional dynamic variables in the form of the distribution and motion of
its own mass.

Own mass is not part of the matter that fills the universe. However, its
presence affects the geometry and thus it interacts with matter through some
form of gravitational force. This effect of the presence of its own mass
means that it is not possible to associate a distinguished frame of
reference with.

\section{PROPER TIME OF THE UNIVERSE}

Above, we have considered a variant of describing the dynamics of the
universe in terms of coordinate time, which can be called external. The
principle of covariance is needed precisely in order to ensure the
independence of physical laws from the choice of this coordinate time. But
this description looks as if there were some external observer for the
universe. Therefore, the description in terms of internal time remains
relevant, which must be introduced in the dynamic interpretation of
solutions of gravitational constraints. There is no single approach to this
issue. The general idea is that the picture of motion in time that is
familiar to us arises in the semiclassical approximation \cite{Giul}. When
analyzing simple mini-superspace models of the universe, one of the
fundamental dynamical variables of the theory is chosen as the time
parameter. It seems natural to choose the $3D$ volume of the universe
(volume logarithm) as the time parameter (see \cite{MTW}). This choice is
consistent with the hyperbolic signature of the constraint Eq.(\ref{10}).
However, here we again encounter the problem of the cosmological
singularity, which in quantum theory takes on the chaotic form of a quantum
billiard \cite{Nicolai}. In a loop QTG near a cosmological singularity, a
massless scalar field is considered \cite{Ash1} as the time parameter.

This section discusses an alternative approach to determining internal time
that does not involve fundamental dynamic variables. It also takes into
account the presence of the own mass of the universe, with which the
selected frame of reference is associated. In classical general relativity,
the frame of reference is given by the lapse and shift functions $N,N_{i}$,
and in the conventional QTG, any dependence on these functions is excluded
as a consequence of the WDW equations. These equations must be modified in
such a way as to take into account the presence of a dedicated frame of
reference. Such a variant of constraints is provided by a construction based
on the Witten identity in the theory of gravity \cite{Witt},\cite{Fadd},
which can be written in integral form for the case of a closed universe \cite%
{GorLukTMF}:

\begin{eqnarray}
&&\left( \chi ,\widehat{W}\eta \right) -\int_{\Sigma }d^{3}x\sqrt{\det g_{ik}%
}\left[ N\left( \chi ,\eta \right) \mathit{H}\right.  \notag \\
&&\left. +N_{i}\left( \chi ,\eta \right) \mathit{H}^{i}\right]  \notag \\
&=&0,  \label{50}
\end{eqnarray}%
where $\chi ,\eta $ are the Dirac bi-spinors on the spatial section $\Sigma $%
. Parentheses denote the scalar product in the space of bi-spinor fields:

\begin{equation}
\left( \chi ,\eta \right) =\int_{\Sigma }d^{3}x\sqrt{\det g_{ik}}\chi
^{+}\eta ,  \label{51}
\end{equation}%
where the cross denotes the Hermitian conjugation (see \cite{GorLukTMF}). If
$\chi =\eta $, identity Eq. (\ref{50}) gives a representation of the
Hamiltonian function of a closed universe in a special gauge \cite{Ash2}, so
that the bi-spinor field $\chi $ completely determines the frame of
reference (see also \cite{Fadd}). It also follows from identity Eq.(\ref{50}%
) that the set of gravitational constraints is equivalent to the operator
equality on the space of bi-spinor fields:

\begin{equation}
\widehat{W}\equiv \widehat{\mathcal{D}}^{2}-\left( -\widehat{\Delta }\right)
=0,  \label{52}
\end{equation}%
where $\widehat{\mathcal{D}}$ is a Hermitian Dirac operator with respect to
the scalar product Eq.(\ref{51}), and $\widehat{\Delta }$ is also a
Hermitian Beltrami-Laplace operator. Note that both operators in this
equality are positive-definite: $\widehat{\mathcal{D}}\eqslantgtr 0,-%
\widehat{\Delta }\eqslantgtr 0$. After quantization, the operator $\widehat{W%
}$ additionally becomes an operator $\widehat{\widehat{W}}$ on the space of
wave functions $\psi \left( g_{ik}\right) $, which is marked with an
additional \textquotedblleft lid\textquotedblright . The "double" operator $%
\widehat{\widehat{W}}$ allows us to write the quantum constraint equations
for the wave function of the universe $\psi \left( g_{ik}\right) $ together
with the conditions for the frame field $\chi $:

\begin{equation}
\widehat{\widehat{W}}\chi \times \psi =0.  \label{53}
\end{equation}%
Thus, in this variant of the quantum theory, the physical states of the
universe should be referred to a selected frame of reference given by the
field $\chi $.

Quantum constraints Eq.(\ref{53}) form a system of differential equations
for the functions $\chi \left( x^{k}\right) $ and $\psi \left(
g_{ik},\varphi _{A}\right) $ (here we have also added matter fields). By
themselves, they do not constitute a specific mathematical problem without
additional conditions. For hyperbolic equations, these are usually the
initial-boundary conditions in the configuration space. Rejecting the
initial-boundary value problem, let us pay attention to another possibility
dictated by the structure of the operator equality Eq.(\ref{52}). It is
obvious that the mean value of any of the two positive operators included in
it has a minimum on the set of solutions of the relation system Eq.(\ref{53}%
). We choose the square of the Dirac operator, which does not explicitly
include matter fields, and therefore, as part of the Hamilton function, can
be called the energy of space. All physical degrees of freedom, including
the transverse components of the gravitational field, are included in the
second positive-definite operator. Thus, we come to the conditional extremum
principle for the space energy

\begin{eqnarray}
\mathit{E} &=&\frac{\left\langle \psi \right\vert \left( \chi ,\widehat{%
\widehat{\mathcal{D}}}^{2}\chi \right) \left\vert \psi \right\rangle }{%
\left\langle \psi \right\vert \left( \chi ,\chi \right) \left\vert \psi
\right\rangle }  \notag \\
&&+\left\langle \left\langle \mathcal{N}\right. \right\vert \left\vert
\left. \widehat{\widehat{W}}\chi \times \psi \right\rangle \right\rangle
\notag \\
&&+\left\langle \left\langle \chi \times \psi \widehat{\widehat{W}}\right.
\right\vert \left\vert \left. \mathcal{N}\right\rangle \right\rangle ,
\label{54}
\end{eqnarray}%
in which the constraint equations Eq.(\ref{53}) are taken into account as
additional conditions with the corresponding Lagrange multipliers $\mathcal{N%
}$. The double brackets here mean the scalar product in the composition
space $\chi \times \psi $. In addition to the minimum value of the space
energy, the conditional extremum problem Eq.(\ref{53}) determines the entire
spectrum of its excitations $\mathit{E}_{\omega }$, which is described by a
certain set of quantum numbers $\omega $. The identification of quantum
numbers $\omega $ with the internal time of the universe is an alternative
to choosing the volume of the universe as a time parameter. An additional
reason for such an interpretation is that such a choice of the cosmological
arrow of time will be consistent with the thermodynamic arrow (entropy), if
the increasing "complexity" of the excited states of space is taken as a
measure of disorder in the universe. However, a large element of
arbitrariness remains in the proposed interpretation until the structure of
the excitation spectrum of space is known.

\section{CONCLUSIONS}

The description of the quantum dynamics of the universe using the wave
functional on the space of world histories $g_{\alpha \beta }\left(
x^{k},t\right) $ allows us to assume that the average values of
gravitational constraints $\left\langle \mathit{H}^{a}\right\rangle $, equal
to zero in classical GR, are nonzero in QTG. We emphasize that here it is
not the constraint operators $\mathit{H}^{a}$ that are subject to averaging,
but the classical expressions Eqs.(\ref{10}), (\ref{11}). The difference
between these averages from zero in quantum theory can arise in the initial
state of the universe. For a new version of the description of dynamics, a
quantum principle of least action is proposed, in which the action operator
is the central object. This operator is obtained by replacing the canonical
momenta in the canonical form of the ADM action by operators of variational
differentiation on the space of wave functionals from world histories.
According to the quantum principle of least action, the evolution of the
universe is described by the eigenfunctions of the action operator. The new
formalism also allows for modification to define the initial state of the
universe. The quantum principle of least action for the "near polar region"
is formulated, in which the action operator is defined on the space of wave
functionals of $4D$ Riemannian metrics in a compact region with one
boundary. In this case, we are interested in the eigenvalue of the action
operator, which is the logarithm of the initial state of the universe. This
state can be called the cosmological vacuum, and there is reason to expect
that it is the source of the universe's own mass. Thus, the appearance of
the universe's own mass is a consequence of quantum theory, and the
associated violation of the covariance of the original theory is an example
of a spontaneous violation of the symmetry of the cosmological vacuum.

\section{ACKNOWLEDGEMENTS}

We are thanks V.A. Franke for useful discussions.

%%%\noindent $^{\ast }$ E-mail address: alex.lukyan@rambler.ru

%%%\noindent $^{+}$ E-mail address: inna.lukyan@mail.ru

%\begin{references}

\end{document}